\documentclass[11pt]{article}
\usepackage[utf8]{inputenc}
\usepackage[margin=1in]{geometry}
\usepackage{cite}
\usepackage{amsmath,amssymb,amsfonts}
\usepackage{algorithmic}
\usepackage{graphicx}
\usepackage{textcomp}
\usepackage{wrapfig}
\usepackage{siunitx}
\usepackage{caption}
\usepackage{subcaption}
\usepackage{xcolor}
\usepackage{soul}
\usepackage{authblk}
\usepackage[hidelinks]{hyperref}

\title{Respiration Rate Sensor Based on Fiber Cavity\\
Attenuated Phase Shift Spectroscopy}

\author[]{Muhammad Fahd Ibrahim\thanks{These authors contributed equally as first authors.}}
\author[]{Shazreen Rashid\textsuperscript{*}}
\author[]{Noor-ul-Amin Nazir}
\author[]{M. Imran Cheema\thanks{Corresponding author: \texttt{imran.cheema@lums.edu.pk}}}

\affil[]{Department of Electrical Engineering, Lahore University of Management Sciences, Pakistan}
\date{}
\usepackage{wrapfig}
\begin{document}
\maketitle

\begin{center}
\section*{Abstract}
\end{center}

\begin{wrapfigure}{r}{0.6\textwidth}
    \vspace{-10pt}
    \centering
    \includegraphics[width=0.6\textwidth]{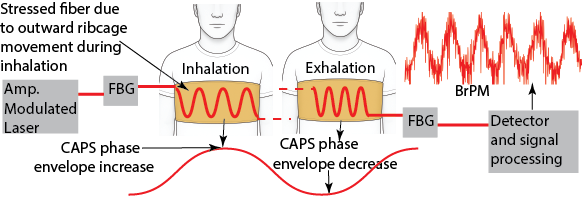}
    \vspace{-10pt}
\end{wrapfigure}

Respiratory rate (RR) is a vital sign with significant diagnostic value. Existing RR
monitors often suffer from baseline drift over time, breaths can be occluded by limb or
body movements, and many systems struggle to resolve shallow or extreme thoracic
motion. Here, we propose a novel RR-monitoring sensor based on fiber-cavity attenuated
phase-shift spectroscopy (CAPS). The sensor comprises a fiber cavity embedded into a
flexible chest binder in a sinusoidal-like pattern worn by the patient. Thoracic expansion
and contraction during breathing modulate the cavity, and RR is extracted through CAPS
measurements. Our sensor exhibits high reproducibility, strong sensitivity to strain and
pressure induced by chest movements, inherent resistance to baseline drift, and the ability
to detect body movements. The system achieves a root-mean-square error of 0.91 breaths
per minute across RR values from 8 to 44 breaths per minute, evaluated on multiple
subjects in various postures. We anticipate that this work will contribute to the development
of comprehensive optical fiber–based sleep-monitoring systems.

\noindent\textbf{Index Terms—}
Cavity attenuated phase-shift spectroscopy (CAPS), optical sensors, fiber cavity, respiration rate monitor, sleep monitor.

\section{Introduction}
\label{sec:introduction}
Real-time, low-cost, and continuous vital signs detection solutions are highly sought after in clinical settings and home-based patient care. Although a wide range of techniques exists for sensing temperature, heart rate, and blood pressure, robust respiration rate (RR) monitoring still requires further work. Previously, researchers have proposed various respiratory monitoring solutions, such as transthoracic impedance plethysmography,
electromyography, and photoplethysmography~\cite{pacela1966impedance, Sinderby1995_EMG, Allen2007_PPG}. However, a common issue in the approaches above is the inaccuracies caused by changes in posture and settling movements.

Electrical impedance tomography has traditionally monitored respiration rate by time-difference reconstruction of lung conductivity using a single excitation frequency~\cite{liu2020textilebelt}. The method provides valuable spatial information on lung ventilation but requires bulky instrumentation and is highly sensitive to motion artifacts, which limits its use outside controlled clinical environments. A recent work in frequency-difference-based impedance tomography utilizes the difference between simultaneous measurements at two frequencies, allowing for the accurate tracking of breathing even during body movement~\cite{ma2025freqdiffEIT}. While this improvement enhances motion tolerance, the technique still relies on multiple electrodes and complex image reconstruction, which constrains its portability and suitability for continuous or wearable use.

Furthermore, ultra-wideband radio-frequency sensors have been explored for contactless monitoring by transmitting pulses from ceiling-mounted devices and reconstructing chest-motion waveforms~\cite{wang2024uwb}. This technique demonstrates strong potential due to its contactless operation and capability to detect signals through clothing and bedding, offering unobtrusive sensing advantages. Nevertheless, existing validations typically consider stationary subjects, and accuracy deteriorates with substantial movement or shifts in postures.

In recent years, there has been a growing interest in alternatives to conventional respiratory monitoring. For example, Yoo et al. proposed a fiber-optic sensor using a thermochromic pigment at the fiber tip to detect temperature differences between inhaled and exhaled air, resulting in a clear breathing waveform~\cite{sec}. This approach is elegant in its simplicity and fully contactless operation; however, it depends on temperature differentials that can be influenced by ambient airflow and humidity, thereby reducing its robustness under real-world conditions.

The aforementioned studies demonstrate significant progress toward non-invasive respiratory monitoring through optical, electrical, and radio-frequency modalities. However, most still face trade-offs between accuracy, motion robustness, and wearability. There remains a need for a compact, flexible, and motion-tolerant sensing approach suitable for continuous use in daily environments.

 Our sensing principle is based on the phase-shift fiber cavity ring-down spectroscopy~(PS-FCRDS) technique, also known as cavity-attenuated phase-shift spectroscopy~(CAPS), which utilizes a Fabry–Pérot cavity formed by two fiber Bragg gratings~(FBGs). The sensing head, positioned between the two FBGs, comprises a series of sinusoidal-like optical‐fiber loops secured onto the surface of a flexible abdomen support belt, effectively creating a sensing belt that lies flush against the wearer’s torso. As the patient inhales and exhales, microstrain in the fiber alters the cavity losses and, consequently, the cavity ring down time, which is correlated to the respiration rate.

 Our system achieved a mean absolute error~(MAE) of 0.91 breaths per minute with respect to the ground truth, across a respiration rate range of 8–44 breaths per minute, evaluated over multiple subjects and postures, including sitting~(with and without back support), supine, prone, and lateral positions. To the best of our knowledge, our work represents the first demonstration of using a Fabry–Pérot cavity for RR monitoring in general and with CAPS in particular.

\begin{figure*}[ht]
   \centering
\includegraphics[width=\linewidth]{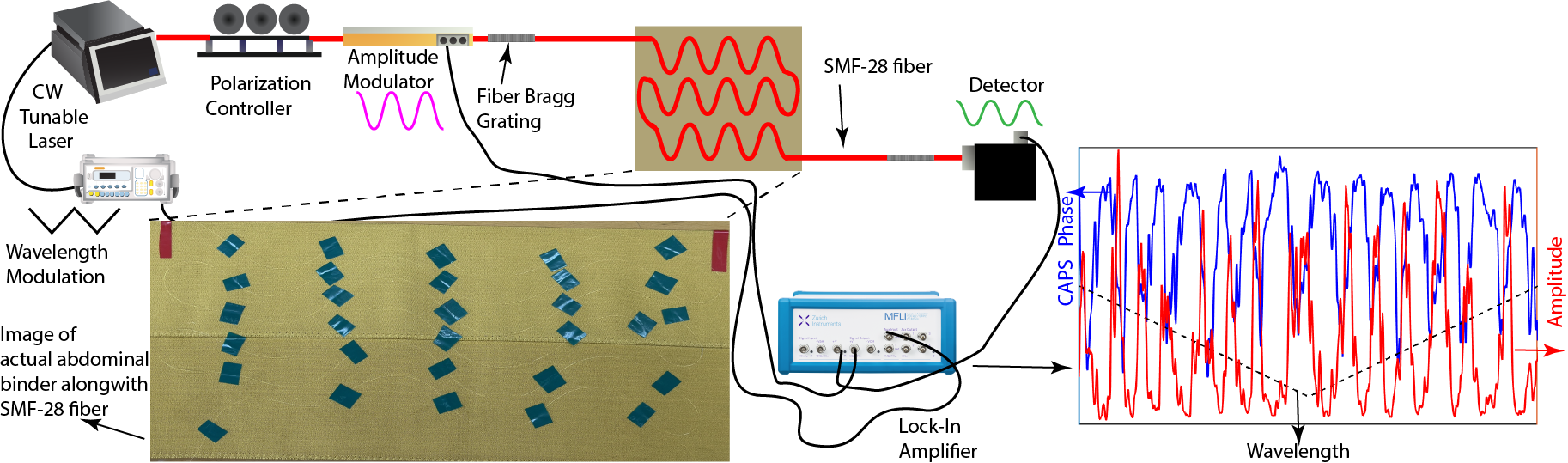}
   \caption{Experimental schematics. A 1550~nm laser is simultaneously wavelength-modulated to scan the fiber cavity resonances and amplitude-modulated for CAPS measurements. Breathing-induced pressure and strain variations on the SMF-28 fiber modulate the cavity response, and the resulting CAPS signals are measured using a lock-in amplifier.}
   \label{fig:schematics}
 \end{figure*}

We now describe the rest of the paper. Section II describes our methodology, including the RR sensor design, its experimental implementation, and the underlying sensing principle. Section III presents the results and details the signal-processing pipeline used to extract the respiration rate from the experimental data, demonstrating robust detection across multiple scenarios (including various seating and supine positions) and diverse body types. Section IV discusses the implications of these findings, addresses limitations, and identifies opportunities for future extensions toward a comprehensive sleep-monitoring system. Finally, Section V concludes the paper by summarizing the key contributions of this work.

\section{Approach}
\label{sec:Approach}
\subsection{Sensor Design}
The sensor schematics are shown in Fig.~\ref{fig:schematics}. Briefly, we use an Eblana Photonics laser diode (EP1550-0-NLW-B26-100FM) driven by a Thorlabs controller~(CLD1015) as the source. An optical cavity is created by splicing two Oeland Inc. Canada FBGs~(R$_1$=86\%, R$_2$=95\%, FWHM: at 1550~nm) at the ends of an SMF-28 fiber. In our setup, the laser is sinusoidally modulated at 2~MHz, 500~mV$_{pp}$ using an external Mach–Zehnder modulator~(Sumitomo Osaka T.MZH1.5-10PD-ADC). To scan the cavity resonances, we apply a continuous triangular current modulation to the laser~(10~Hz, 50~mV$_{pp}$) via a function generator~(Rigol DG1022). The combined current and amplitude modulation enables CAPS measurements at resonance peaks without requiring the conventional PDH locking scheme~\cite{Ghauri2021_AflatoxinM1}. The modulated laser output is coupled into the cavity, while the same signal is simultaneously provided as the reference input to the lock-in amplifier~(Zurich Instruments MFLI). The cavity transmission is detected using an InGaAs photodetector~(Thorlabs DET08CFC/M), with its output connected to the lock-in amplifier.

We attached the silica fiber inside a commercially available abdominal binder (PERFECT brand) in a sinusoidal pattern using PVC tape~(Fig.~\ref{fig:schematics}). Since the fiber is brittle and can break if subjected to sharp bends, the sinusoidal layout provides slack and reduces tension on the fiber during breathing. Multiple waves of fiber were layered along the inner side of the belt to enlarge the sensing area and thus increase sensitivity. We fabricated three belt sizes, large, medium, and small, to accommodate different body types.

\begin{figure}[ht]
   \centering
\includegraphics[width=4in]{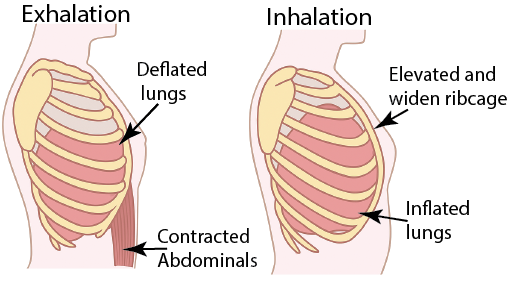}
   \caption{Contraction and expansion of the thoracic cage during exhalation and inhalation, respectively. (Image adapted from~\cite{guyton2016physiology})}
   \label{fig:ex_inhale}
 \end{figure}

During inhalation, the anteroposterior, vertical, and transverse (side-to-side) diameters of the thoracic cavity and chest wall increase due to the coordinated contraction of the diaphragm and external intercostal muscles, as shown in Fig.~\ref{fig:ex_inhale}~\cite{guyton2016physiology}.
Informed by this physiological mechanism, the fiber layout was designed to ensure that, when worn, the sensing cavity contacts only the ventral and lateral regions of the torso, targeting areas that exhibit the most significant rib displacement during respiration. The upper side of the chest, located just below the armpit, was intentionally left uncovered to minimize motion artifacts arising from arm movements.

\subsection{CAPS Sensing Principle with Fiber Cavity}

When a CW laser light is sinusoidally modulated at an angular frequency, $\omega$, and injected into a Fabry-Pérot fiber cavity, a phase shift, $\phi$, is introduced into the light at the cavity output, given by~\cite{Ghauri2021_AflatoxinM1}:
\begin{equation}\label{eq:phi}
\tan\phi = -\omega\tau,
\end{equation}
where $\tau$ is the cavity ring down time. If the cavity length and effective refractive index are $L$ and $n$, respectively then the cavity round-trip time, $t_{rt}$, is given by
\begin{equation}\label{eq:rt}
t_{rt}=\frac{2nL}{c},
\end{equation}
where $c$ is the speed of light in a vacuum. Assuming that $\alpha_{\mathrm{rt}}$ represents all cavity losses, then the cavity ring down time $\tau$ can be written as:
\begin{equation}\label{eq:tau}
      \tau = \frac{t_{\mathrm{rt}}}{\alpha_{\mathrm{rt}}}.
    \end{equation}
When a person inhales, the thoracic and abdominal cavities both expand. This expansion causes the optical fiber that makes up the cavity to be stretched by a strain, leading to a change in the cavity length, given by:
\begin{equation}\label{eq:deltaL}
\Delta L = \varepsilon\,L
\end{equation}
where $\varepsilon$ is the axial strain on the fiber. Owing to the way the fiber is in flush contact with the thorax and abdomen, a transverse pressure $p$ is also applied on the fiber, which induces additional losses in the cavity. Overall, both $\varepsilon$ and $p$ will impact $\tau$ during the breathing process. The CAPS phase change can be deduced by differentiating equation~\eqref{eq:phi}:
\begin{equation}\label{eq:dphi_dtau}
  d\phi \;=\; \frac{d\phi}{d\tau}\,d\tau
  \;=\; -\frac{\omega}{1+(\omega\tau)^2}\,d\tau.
\end{equation}
We expand $d\tau$ in the two inputs $\varepsilon$ and $p$:
\begin{equation}\label{eq:dtau_dp_de}
  d\tau \;=\; \frac{\partial\tau}{\partial\varepsilon}\,d\varepsilon
  + \frac{\partial\tau}{\partial p}\,dp.
\end{equation}

Using equation~\eqref{eq:tau}, the partial derivative with respect
to a generic perturbation \(x\in\{\varepsilon,p\}\) is
\begin{equation}\label{eq:dtau_dx_general}
  \frac{\partial\tau}{\partial x}
  \;=\;
  \frac{1}{\alpha_{\mathrm{rt}}}\,\frac{\partial t_{\mathrm{rt}}}{\partial x}
  \;-\;
  \frac{t_{\mathrm{rt}}}{\alpha_{\mathrm{rt}}^2}\,\frac{\partial \alpha_{\mathrm{rt}}}{\partial x}.
\end{equation}
Differentiating the round-trip time equation~\eqref{eq:rt} explicitly:
\begin{equation}\label{eq:dtrt_dx}
  \frac{\partial t_{\mathrm{rt}}}{\partial x}
  \;=\; t_{\mathrm{rt}}\Big(\frac{1}{n}\frac{\partial n}{\partial x} + \frac{1}{L}\frac{\partial L}{\partial x}\Big).
\end{equation}

For the two variables we have:
\begin{itemize}
  \item \textbf{Axial strain} $\varepsilon$: $\dfrac{\partial L}{\partial\varepsilon}=L$, hence
    \begin{equation}\label{eq:dtrt_de}
      \frac{\partial t_{\mathrm{rt}}}{\partial\varepsilon}
      \;=\; t_{\mathrm{rt}}\Big(1+\frac{1}{n}\frac{\partial n}{\partial\varepsilon}\Big).
    \end{equation}
  \item \textbf{Transverse pressure} $p$: to first order we neglect global length change due to $p$,
    so $\dfrac{\partial L}{\partial p}\approx 0$, and
    \begin{equation}\label{eq:dtrt_dp}
      \frac{\partial t_{\mathrm{rt}}}{\partial p}
      \;=\; t_{\mathrm{rt}}\Big(\frac{1}{n}\frac{\partial n}{\partial p}\Big).
    \end{equation}
\end{itemize}

By substituting equation~\eqref{eq:dtrt_de} and equation~\eqref{eq:dtrt_dp} into equation~\eqref{eq:dtau_dx_general}, we obtain
\begin{align}
  \frac{\partial\tau}{\partial\varepsilon}
  &= \frac{1}{\alpha_{\mathrm{rt}}}\,t_{\mathrm{rt}}\Big(1+\frac{1}{n}\frac{\partial n}{\partial\varepsilon}\Big)
    - \frac{t_{\mathrm{rt}}}{\alpha_{\mathrm{rt}}^2}\,\frac{\partial\alpha_{\mathrm{rt}}}{\partial\varepsilon},
    \label{eq:dtau_de_full}\\[6pt]
  \frac{\partial\tau}{\partial p}
  &= \frac{1}{\alpha_{\mathrm{rt}}}\,t_{\mathrm{rt}}\Big(\frac{1}{n}\frac{\partial n}{\partial p}\Big)
    - \frac{t_{\mathrm{rt}}}{\alpha_{\mathrm{rt}}^2}\,\frac{\partial\alpha_{\mathrm{rt}}}{\partial p}.
    \label{eq:dtau_dp_full}
\end{align}

Inserting equation~\eqref{eq:dtau_de_full} and equation~\eqref{eq:dtau_dp_full} into equation~\eqref{eq:dphi_dtau}, we obtain:

\begin{equation}\label{eq:dphi_de}
  \boxed{%
    \displaystyle
    \frac{\partial\phi}{\partial\varepsilon}
    \;=\;
    -\frac{\omega\tau}{1+(\omega\tau)^2}
    \left[
      1+\frac{1}{n}\frac{\partial n}{\partial\varepsilon}
      -\frac{1}{\alpha_{\mathrm{rt}}}\frac{\partial\alpha_{\mathrm{rt}}}{\partial\varepsilon}
    \right]
  }
\end{equation}

\begin{equation}\label{eq:dphi_dp}
  \boxed{%
    \displaystyle
    \frac{\partial\phi}{\partial p}
    \;=\;
    -\frac{\omega\tau}{1+(\omega\tau)^2}
    \left[
      \frac{1}{n}\frac{\partial n}{\partial p}
      -\frac{1}{\alpha_{\mathrm{rt}}}\frac{\partial\alpha_{\mathrm{rt}}}{\partial p}
    \right]
  }
\end{equation}

 It is well-established that applying either strain or pressure induces losses in optical fibers~\cite{Nagano1978_RefractiveIndex,Schermer2007_BendLoss,Widiyatmoko2022_Macrobending}; therefore, both $\frac{\partial \alpha_{\mathrm{rt}}}{\partial \varepsilon}$ and $\frac{\partial \alpha_{\mathrm{rt}}}{\partial p}$ are positive.   In the context of our sensing system, equations~\ref{eq:dphi_de}-\ref{eq:dphi_dp} indicate that during inhalation, when pressure and strain rise, the CAPS phase decreases, whereas during exhalation, when pressure and strain fall, the CAPS phase increases.

\subsection{Respiratory Reference Using Patient Monitor}
\begin{figure}[!t]
   \centering
   \includegraphics[width=\linewidth,height=1.5in,keepaspectratio]{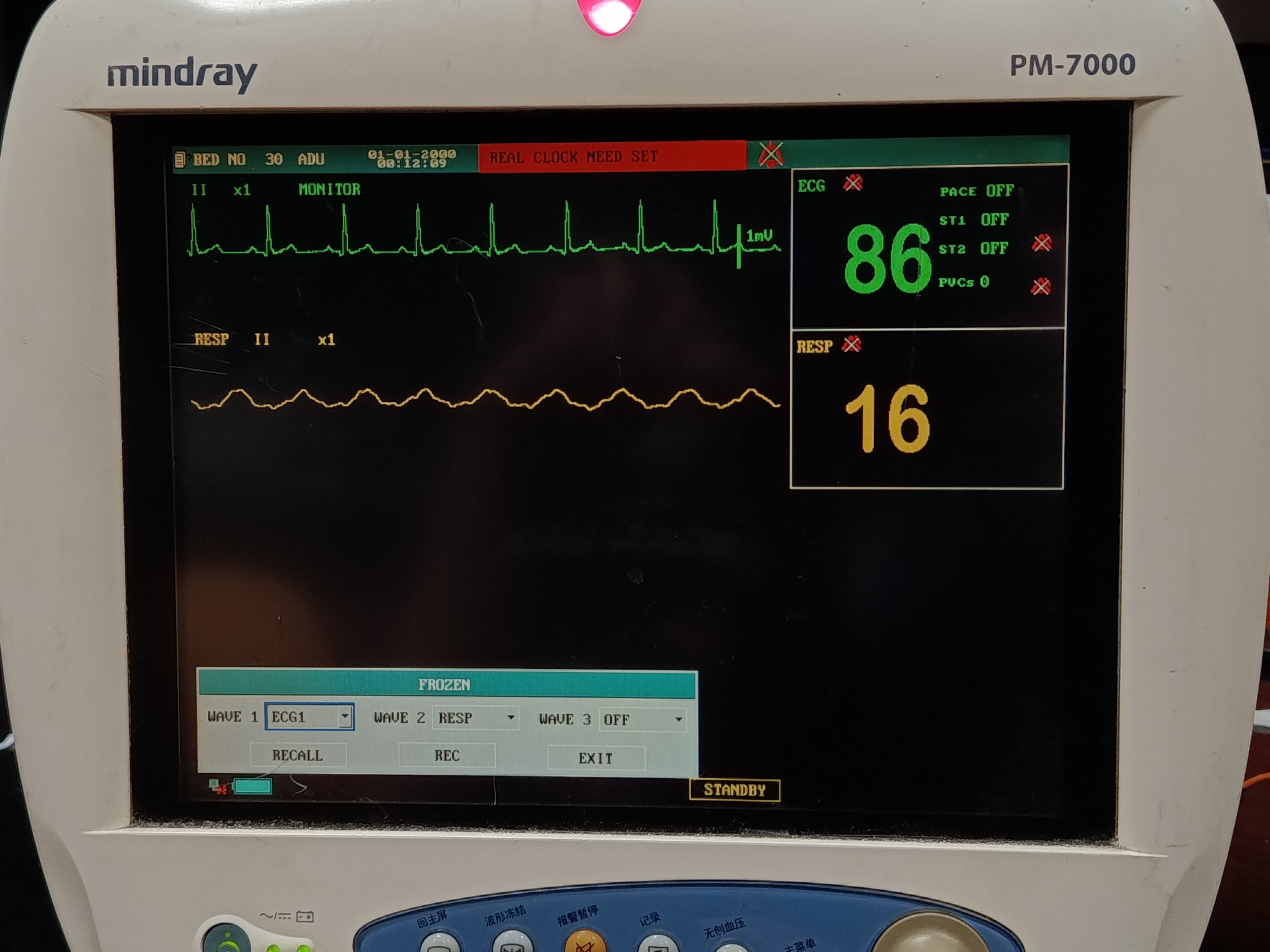}
   \caption{The Mindray monitor used as a reference device for respiration monitoring.}
   \label{fig:mindray}
 \end{figure}

\color{black}
To establish an RR reference, we use a commercially available patient monitor, Mindray PM-7000, as shown in Fig.~\ref{fig:mindray}. It has a one breath per minute (BrPM) resolution, with a defined precision of "$\pm$2 BrPM or $\pm$2\%, whichever is greater" for 7-150 BrPM. The measurement technique is thoracic impedance, with a bandwidth of 0.2~Hz to 2~Hz and a measurement range (for adults) of 0-120~BrPM. The monitor's three probes are placed at their respective positions~(LL, LA, RA), as per the monitor manual, and then the volunteers wear the cavity jacket and lie down on their backs.

\section{Results}
\label{sec:Results}

\begin{figure}[!t]
   \centering
\includegraphics[width=0.8\columnwidth]{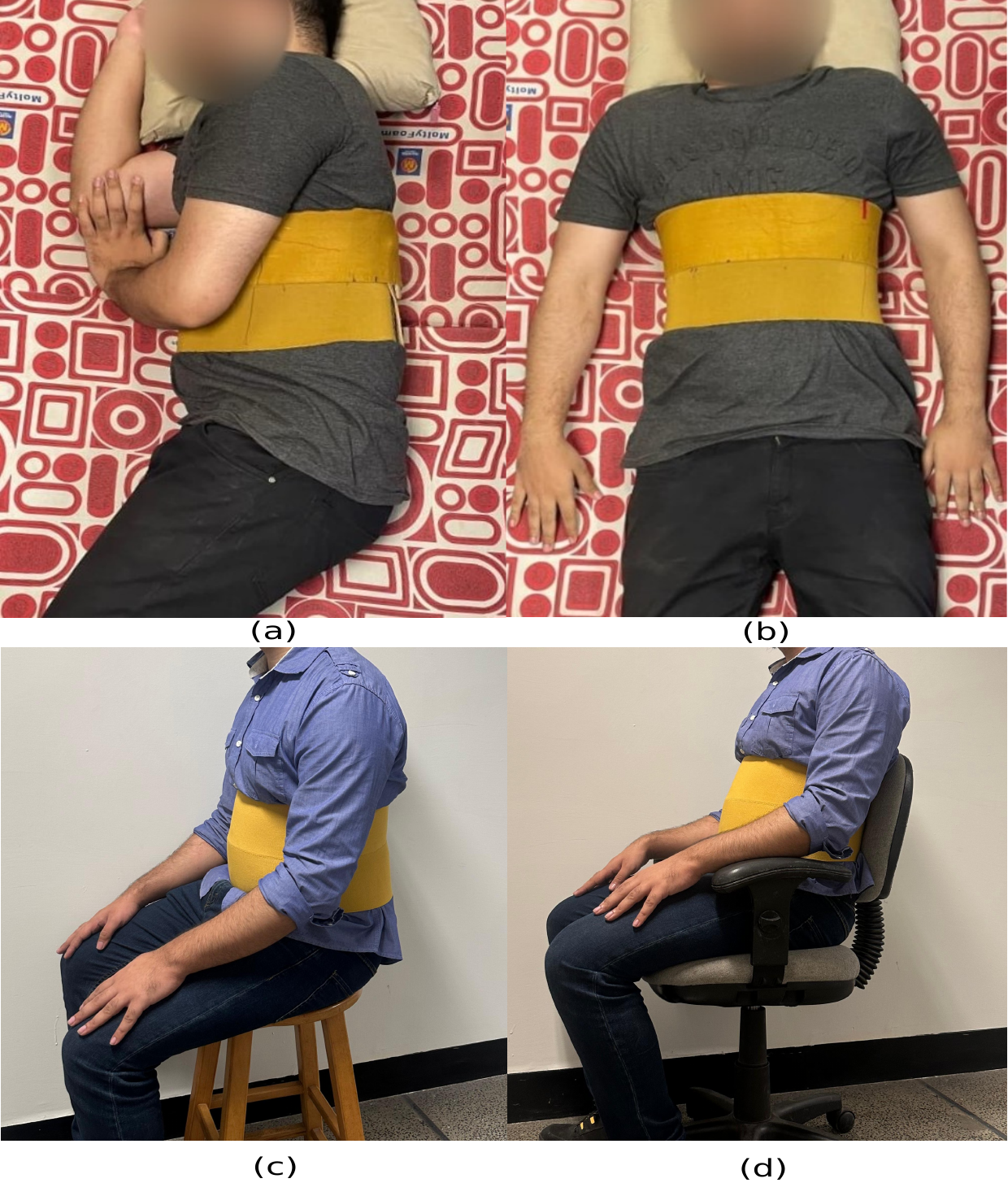}
   \caption{A volunteer in different representative postures.
   (a) lying on the side, (b) lying supine,
(c) sitting without back support, and
(d) sitting with back support.}
   \label{fig:positions}
\end{figure}

We evaluate the sensor performance using data from four volunteers, whose characteristics are summarized in Table~\ref{tab:subjects}.
\begin{table}[!t]
\centering
\caption{Characteristics of volunteers tested with the proposed sensor}
\label{tab:subjects}
\begin{tabular}{c c c c}
\hline
\textbf{Volunteer} & \textbf{Height (cm)} & \textbf{Weight (kg)} & \textbf{BMI (kg/m$^2$)} \\
\hline
1 & 167.64 & 70    & 24.9 \\
2 & 185.42 & 107.5 & 31.3 \\
3 & 185.42 & 63    & 18.3 \\
4 & 177.80 & 89    & 28.2 \\
\hline
\end{tabular}
\end{table}
To test the sensor's robustness, we include diverse participants with different body types, weights, and heights, and record CAPS measurements in multiple positions: sitting without back support, sitting with back support, supine, lying on the right side, lying on the left side, and standing, as shown in Fig.~\ref{fig:positions}.

\begin{figure*}[!t]
  \centering
  \captionsetup[subfigure]{justification=centering} 

  \begin{subfigure}{0.48\textwidth}
    \centering
    \includegraphics[width=\linewidth]{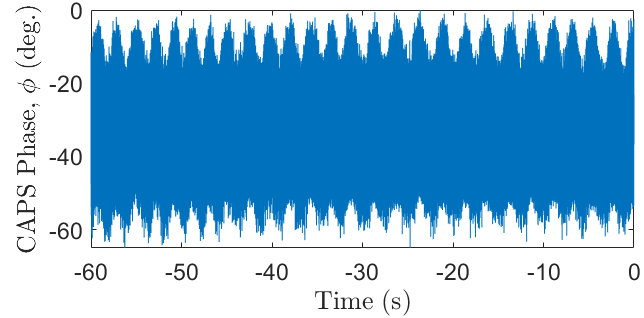}
    \caption{}
    \label{fig:wave_a}
  \end{subfigure}\hfill
  \begin{subfigure}{0.48\textwidth}
    \centering
    \includegraphics[width=\linewidth]{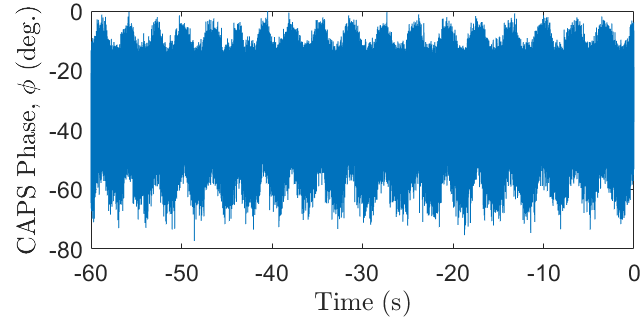}
    \caption{}
    \label{fig:wave_b}
  \end{subfigure}

  \vspace{6pt}

  \begin{subfigure}{0.48\textwidth}
    \centering
    \includegraphics[width=\linewidth]{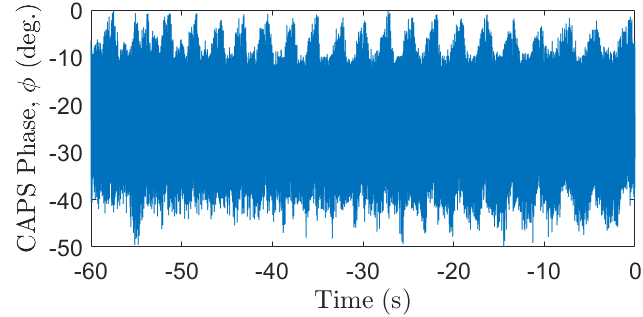}
    \caption{}
    \label{fig:wave_c}
  \end{subfigure}\hfill
  \begin{subfigure}{0.48\textwidth}
    \centering
    \includegraphics[width=\linewidth]{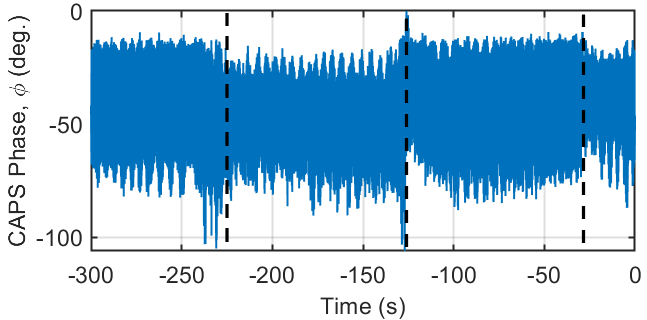}
    \caption{}
    \label{fig:wave_d}
  \end{subfigure}
  \caption{Representative CAPS phase waveforms measured from subjects in different postures:
(a) standing,
(b) sitting with back support,
(c) sitting without back support,
(d) lying down. In (d), the subject is initially supine; after the first dotted line, the subject turns to one side; after the second dotted line, to the opposite side; and finally returns to the supine position.}
  \label{fig:waveforms}
\end{figure*}

The CAPS measurements for each tested posture are shown in Fig.~\ref{fig:waveforms}. The phase signal was offset such that the baseline value of the signal was set to zero degrees. This was done to enhance visual clarity and did not affect any underlying measurements. Despite differences in body postures, the data collected from multiple volunteers exhibit similar signal characteristics. The variation in breathing patterns with posture is consistent with prior reports on the influence of body position on respiratory mechanics~\cite{Decker2024_BodyPositionBreathing}.

\begin{figure}[!t]            
  \centering
  \begin{subfigure}{\linewidth}
    \centering
    \includegraphics[width=\linewidth]{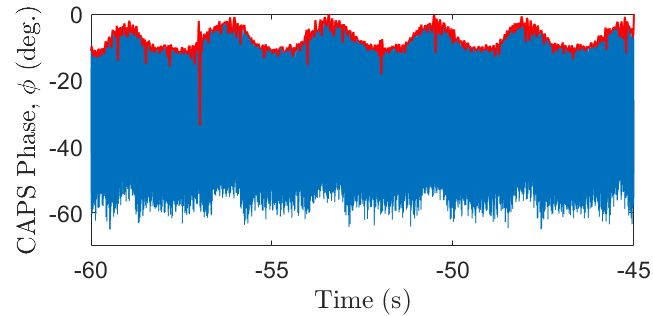}
    \caption{}
    \label{fig:suba}
  \end{subfigure}

  \vspace{3pt}

  \begin{subfigure}{\linewidth}
    \centering
    \includegraphics[width=\linewidth]{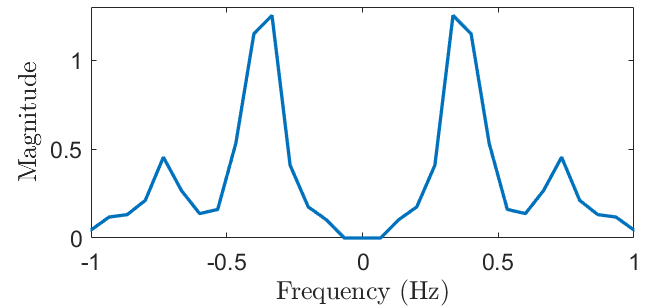}
    \caption{}
    \label{fig:subb}
  \end{subfigure}

  \caption{A segment of the phase signal spanning 15 seconds with its fitted envelope is shown in (a). The corresponding result of applying FFT to the envelope is shown in (b), indicating 20 BrPM.}
  \label{fig:two_stacked}
\end{figure}

We process all our CAPS data signals in MATLAB. The breathing signal is extracted by determining the envelope of the CAPS waveform, as shown in Fig.~\ref{fig:suba}. The upper and lower envelopes of the CAPS waveform are determined by forming the waveform's analytic signal via the Hilbert transform, which is implemented through the discrete Fourier transform~\cite{Luo2009_HilbertTransformEngineeringApplications}. Fig.~\ref{fig:suba} shows the upper envelope, which effectively captures the breathing waveform. To determine the breathing rate, we apply the Fast Fourier Transform (FFT) to this envelope. The FFT transforms the time-domain envelope into its frequency-domain representation, producing a spectrum~(Fig.~\ref{fig:subb}) that reveals the distribution of power across different frequencies~\cite{Oppenheim1997_DSP}. The fundamental frequency of respiration is identified as the dominant peak in this spectrum, corresponding to the breathing rate in Hz (later converted to breaths per minute). We perform the analysis by segmenting the signal into consecutive 15-second windows, and the breathing rate reported each minute is obtained by averaging the results from four such 15-second windows. This approach improves robustness, since respiration rate naturally varies within a minute~\cite{vandenBosch2021_BreathingVariability}.

\begin{figure}[!t]
   \centering
   \includegraphics[width=\linewidth]{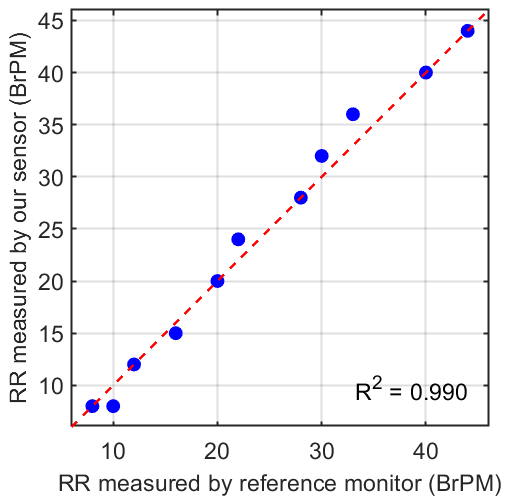}
   \caption{Correlation scatter plot of respiratory rates measured by our CAPS sensor and the reference device, showing the correlation coefficient. RR: respiratory rate}
   \label{fig:scatter}
 \end{figure}

Our sensing results show a root-mean-square error (RMSE) of 0.91 BrPM and a coefficient of determination (R²) of 0.99, as illustrated in Fig.~\ref{fig:scatter}. This performance demonstrates strong agreement with the gold-standard Mindray PM-7000 reference monitor. To evaluate the limits of our system, we successfully measured respiration rates up to 60 breaths/min while maintaining comparable accuracy across the extended range. The CAPS waveforms are also in line with our derived equations~\ref{eq:dphi_de}-\ref{eq:dphi_dp}, as during inhalation, when pressure and strain are at their highest, we obtain the most positive values in the CAPS waveform, and vice versa for exhalation.

\section{Discussion}
\label{sec:Disucssion}
Our sensor maintains accurate breathing rate measurements even during substantial posture changes, as shown in Fig.~\ref{fig:wave_d}, where a volunteer changes positions over a brief interval. As illustrated in Fig.~\ref{fig:wave_d}, each body movement corresponds to a shift in the minimum and maximum of the phase signal. This occurs because baseline stress and strain on the fiber change with posture. By calculating the difference between the minima and maxima in each time window and comparing these across consecutive windows, we successfully detect the occurrence of movements~\cite{Ibrahim2023_LMM_Healthy}.

\begin{figure}[!t]
   \centering
   \includegraphics[width=\linewidth]{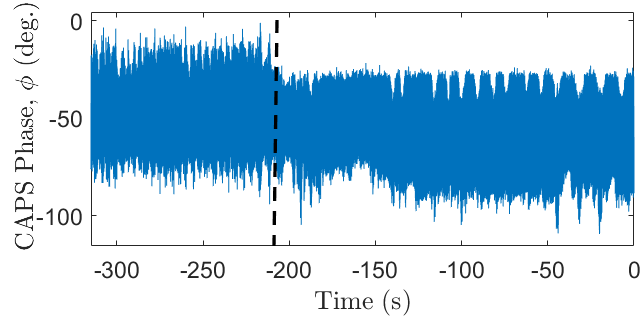}
    \caption{A segment of the CAPS phase measured from a subject during sleep. The dashed vertical line indicates a body movement event.}
    \label{fig:wave_e}
\end{figure}

We also evaluate the sensor during sleep to eliminate potential conscious bias of the subject and to test performance under natural conditions. The results are shown in Fig.~\ref{fig:wave_e}, demonstrating the potential of this methodology for sleep quality monitoring applications. Future studies could extend our work toward developing a comprehensive sleep-quality monitoring system for various applications, including assessing the sleep quality of clinically depressed patients who generally exhibit excessive body movements during their sleep. In addition to reliable respiration-rate detection, our sensor could be enhanced to automatically characterize sleep posture transitions, identify abnormal respiratory events such as apnea or hypoventilation, and correlate respiratory patterns with sleep quality indices. Moreover, integrating machine learning–based classification of CAPS phase irregularities and long-term testing of our sensor could enable early screening for sleep-related breathing disorders, providing a low-cost and user-friendly alternative to conventional polysomnography.

\section{Conclusion}
This paper demonstrates the first application of respiration monitoring using CAPS. We embed a portion of the fiber cavity inside a chest binder in a sinusoidal-like pattern to target the parts of the thoracic cavity that exhibit the most movement during respiration. The stress and strain effects on the fiber cavity during the inhaling and exhaling process bring information about the breathing rate, which is accurately detected by our CAPS sensor. Most major respiration monitoring techniques suffer from posture-dependent efficacy, while our sensor provides accurate measurements in various postures, even while lying down and sleeping. We anticipate that our demonstrated sensor can lead to an accurate and robust solution for comprehensive sleep monitoring for various mental health applications.

\label{sec:Conclusion}

\bibliographystyle{IEEEtran}
\bibliography{refrences}

\end{document}